# EQUATION OF STATE OF URANIUM AND PLUTONIUM

*Dalton Ellery Girão Barroso*
*Military Institute of Engineering*
*Department of Nuclear Energy*[*]
*Email: Dalton@ime.eb.br*

## ABSTRACT

*The objective of this work is to define the parameters of the three-term equation of state for uranium and plutonium, appropriate for conditions in which these materials are subjected to strong shock compressions, as in cylindrical and spherical implosions.*

*The three-term equation of state takes into account the three components of the pressure that resist to compression in the solid: the elastic or "cold" pressure (coulombian repulsion between atoms), the thermal pressure due to vibratory motion of atoms in the lattice of the solid and the thermal pressure of electrons thermally excited. The equation of state defined here permits also to take into account the variation of the specific heat with the transition of the solid to the liquid or gaseous state due to continued growth of temperature in strong shock compressions.*

*In the definition of uranium equation of state, experimental data on the uranium compression, available in the open scientific literature, are used. In the plutonium case, this element was considered initially in the alpha-phase or stabilized in the delta-phase. In the last case, an abrupt and instantaneous transition to the alpha-phase was considered when the delta-phase plutonium is submitted to strong compressions. From the alpha-phase and subsequent phases characteristics of the plutonium under compression, we attempt to demonstrate the possible validity of applying the same equation of state defined here for uranium to the plutonium, especially in higher pressure ranges of interest corresponding to high densities, when both elements transit to the same allotropic solid phase and, subsequently (or directly, depending on the temperature value), to the liquid state.*

*Finally, we tried to simulate numerically the atomic bomb "Fat Man" (detonated at Trinity and Nagasaki), using in the plutonium implosion and compression the equation of state defined here for this element.*

# CONTENTS

# 1. Introduction

In simulation of dynamic compression of materials, caused by the propagation of strong shock waves, it is necessary to define appropriate equations of state for materials subjected to various conditions of density, temperature and pressure. An equation of state comprises a relation among pressure, density (or specific volume) and temperature, $P=P(\rho,T)$, although in some cases, when the temperature determination is not relevant, it is convenient to define the equation of state as a function of density and, directly, the specific internal energy, $P=P(\rho,E)$, simplifying the thermodynamic calculation.

The so-called "three-term equation of state",[1,2] defined in terms of density and temperature, takes into account the three components of the pressure that resist to compression in the solid: the elastic (or "cold") pressure, $P_c$ (Coulombian repulsion between atoms); the thermal pressure, $P_a$, due to vibratory motion of atoms in the lattice of the solid; and the thermal pressure of electrons, $P_e$, thermally excited. This equation, which may also take into account variations in the specific heat and in the Gruneisen coefficient with the density and temperature,[2] is defined in section 2.

Setting appropriate equations of state for the fissile materials uranium and plutonium is of fundamental importance in nuclear explosives analysis, since in the implosion method used to assemble the supercritical fissile mass, these materials are submitted to strong shock compressions caused by the detonation of the chemical high-explosive used in the process of implosion.[3] Densities two or more times higher than the normal density of uranium and plutonium, pressures of the order of tens of megabars (1 Mbar=$10^6$ atm) and temperatures above $10^5$ K can be achieved in the center of the fissile mass during the process of implosion. It is therefore crucial the thermodynamic description of the fissile materials under such conditions and the definition of appropriate equations of state.

Static and dynamic compression data of uranium for a wide range of pressure and density are available in the open scientific literature[4-10]. However, for the plutonium, no unambiguous experimental compression data exist for this element, perhaps because plutonium is radioactive, toxic and subjected to strict safeguards. This greatly complicates the definition of an equation of state for the plutonium, mainly because it is considered one of the most complex elements of nature.[11-13]

Plutonium has six allotropic phases at normal pressure and undergoes transitions to different phases under compression. This last fact is very important, considering that the final solid phase acquired by the plutonium under high static compression (body-centered cubic phase - see section 5) is similar to that acquired by the uranium. This fact makes legitimate, under such circumstances, to extent the use of the uranium equation of state to the plutonium, since these two elements have atomic densities and volumes almost equal.

The plutonium phase with lowest density is the delta-phase, with theoretical density of 15.9 g/cm$^3$. Although existing only in the range temperature of 310-452$^0$C, it can be stabilized at room temperature adding to plutonium, in small quantities ($\cong$1% in weight), elements such as gallium or aluminum. It is commonly accepted that plutonium used in nuclear explosive is stabilized in its delta-phase (with the addition of gallium since with this element there is no undesirable production of neutrons from the ($\alpha$, n) reaction), mainly because in this phase plutonium has better metal properties than in the alpha-phase — hard, brittle and more difficult to shape.

Apparently disadvantageous from the viewpoint of criticality, because of its lower density, the plutonium stabilized in the delta-phase actually undergoes a sudden transformation to the alpha-phase (of higher density) when submitted to strong shock compression. In fact, this occurs at pressures below 10 kilobars,[14-16] well below the characteristic pressures generated by chemical high-explosive. This sudden change of phase (a phenomenon that occurs in many other materials) partly compensates this disadvantage since, under the same shock pressures, plutonium compression from its delta-phase reaches densities not so far from those that would be achieved if it were compressed from its alpha-phase, precisely due to this sudden change of phase. An additional advantage of the plutonium in the delta-phase is that it is possible to obtain a more subcritical mass in that phase, whose critical mass (not reflected) is about 16 kg, well above the critical mass of plutonium in the alpha-phase ($\cong$10 kg for pure Pu-239).

Therefore, in the definition of equations of state, it is necessary to establish a model that represents a sudden phase change that occurs in certain materials submitted to compression. In the case of plutonium, the situation is simplified by the fact that the change of delta-phase to the alpha-phase occurs at pressure much lower, as we already mentioned, than the pressures produced by chemical high-explosives. We can even consider them negligible (kept equal to zero) during the time that the material takes to reach the density corresponding to the alpha-phase (this is a situation analogous to the compressing of a porous mass of plutonium with a density equal to the density of plutonium in delta-phase - see section 3). The phase transition is therefore considered to occur instantaneously without any resistance of delta-phase plutonium to compression (thus eliminating the need to establish an equation of state for that phase).

## 2. The three-term equation of state

As we mentioned in section 1, at extremely high shock pressures it is necessary to consider the three components of the pressure that resist to compression in the solid: the elastic pressure (Coulombian repulsion between atoms), $P_c$, the thermal pressure due to vibration of atoms, $P_a$, and the thermal pressure of electrons (thermally excited with the extreme increase of temperature), $P_e$. These three components are present below (in consecutive order) in the equation of state used in this work and formulated by Kormer et al.:[2]

$$P(\rho,T) = P_c(\rho) + \gamma'_g(\rho,T)\rho E_a(\rho,T) + \gamma_e \rho E_e(\rho,T). \tag{1}$$

The specific internal energy is given by:

$$E(\rho,T) = E_c(\rho) + E_a(\rho,T) + E_e(\rho,T), \tag{2}$$

where $E_a$ and $E_e$ are the components of the thermal energy due to atoms and electrons, respectively, and expressed by:

$$E_a(\rho,T) = \frac{3[2+Z(\rho,T)]}{2[1+Z(\rho,T)]} R(T-T_0) + E_0, \tag{3}$$

$$E_e(\rho,T) = \frac{b^2}{\beta(\rho)} \ln\{\cosh[\beta(\rho)T/b]\}. \tag{4}$$

$\gamma'_g$, $\beta$ e $\gamma_g$ are given by:

$$\gamma'_g(\rho,T) = \frac{2[3\gamma_g(\rho) + Z(\rho,T)]}{3[2+Z(\rho,T)]}, \tag{5}$$

$$\beta(\rho) = \beta_0(\rho_0/\rho)^{\gamma_e}, \tag{6}$$

$$\gamma_g(V) = \gamma_{g0}V/V_0 + 2/3(1-V/V_0); \quad (V=1/\rho). \tag{7}$$

In the above expression, R is the universal gas constant (divided by the atomic mass number: R=Nk, where $N = N^{Avogadro}/\overline{A}$ is the specific atomic density); $\gamma_g$, the Gruneisen coefficient, is given by a well-known empirical expression with extrapolation to the value of 2/3 (ideal gas) when $V \to 0$; $\beta$ is the coefficient of the electronic specific heat; $\gamma_e$ is called electronic Gruneisen coefficient (often put equal to 0.5, near the value calculated by the Thomas-Fermi theory for the degenerate free electrons gas in the solid,[17,1] although its determination is controversial and may, for some metals, deviate quite of the above value[18]); b is a constant; $E_0$ and $T_0$ are, respectively, the initial specific internal energy and temperature.

The variable $Z(\rho,T)$ has been introduced to take into account variations in the specific heat and in the Gruneisen coefficient with the density and temperature. Its value is calculated by the expression:

$$Z(\rho,T) = \ell RT/c_c^2 . \qquad (8)$$

where $\ell$ is a constant characteristic of the material and $c_c^2 = dP_c/d\rho$. The value of Z is proportional to the ratio of the thermal and elastic components of pressure in the solid.

The specific heat of the lattice is given by the expression:

$$c_v(\rho,T) = \partial E_a/\partial T = \frac{3}{2}R\frac{1}{(1+Z)}[(2+Z) - \frac{Z(1-T_0/T)}{(1+Z)}] . \qquad (9)$$

The eq.(9) represents an interpolation: when Z=0, $c_v$=3R (or equal to $c_{v0}$), the characteristic value of the solid state, and when Z→∞, $c_v$=3/2R, corresponding to the gaseous state.

The variation of $\gamma'_g$ (the new Gruneisen coefficient set as a function of density and temperature) takes into account the influence of the inharmonious vibratory motion of atoms with the temperature increase. Of course, when Z=0, $\gamma'_g = \gamma_g(V)$; when $Z \Rightarrow \infty, \gamma'_g \Rightarrow$ 2/3, the value of the ratio between pressure and energy density for an ideal gas.

The expression (4) considers the influence of the temperature on the calculation of energy of electrons in the degenerate state, i.e., the expression is an interpolation which includes results of the Thomas-Fermi theory for T≠0. This correction becomes important for temperatures above 30-50x10³ K. For lower temperatures, the expression (4) reduces to the usual and simpler expression for a fully degenerate electron gas:[1]

$$E_e = \beta(\rho)T^2/2. \qquad (10)$$

The elastic (cold) pressure is given by the Altshuler expression:[19]

$$P_c(\delta) = Q\{\delta^{2/3}\exp[q(1-\delta^{-1/3})] - \delta^{4/3}\}, \qquad (11)$$

and the elastic energy, $E_c = -\int_{V_0}^{V} P_c(V)dV$, by the expression:

$$E_c(\delta) = (3Q/\rho_{0k})\{q^{-1}[\exp(q(1-\delta^{-1/3}))-1]+(1-\delta^{1/3})\}, \tag{12}$$

where $\rho_{0k}$ is the density at 0 Kelvin ($\rho_{0k} \cong \rho_0$), $\delta = \rho/\rho_{0k}$, Q e q are constants to be determined.

Thus, the parameters to be determined in the three-term equation of state, for each material, are: $\gamma_{g0}, \gamma_e, \beta_0, \ell, b, q$ and $Q$.

## 3. Allotropic variations in materials under compression

It is well known the property that certain crystalline solid materials have to undergo sudden changes in the rearrangement of their atoms, i.e., sudden changes in crystalline phase (from a lower density one to another higher density), when subjected to certain shock pressures;[1,4,20] for example, the transition phase (bcc-hcp) in iron submitted to dynamic pressures greater than 130 kbar. To these phase transitions are associated two unusual phenomena: the spontaneous division of the shock wave into two shock waves with different propagation speeds (depending on the pressure applied) and the possibility of rarefaction shock (see the cited references).

Figure 1 shows the typical Hugoniot curve of a material with phase transformation. $P_I$ is the Hugoniot curve for the material in phase 1 and $P_{II}$, the Hugoniot curve for the material in phase 2. Between points A and B coexist the two phases. Regarding the points marked in the figure, we observe the following: If $P<P_A$ there is a normal shock wave propagating in the material without phase change (and whose speed is determined by the slope of the segment joining the initial point O and the point on the curve); if $P>P_C$, again there is a single shock wave, but with abrupt change of phase occurring in the wavefront; if $P_C>P>P_A$, the slope of the segment 0-N is lower than the slope of segment O-A, causing such anomalous situation the formation of two shock waves in the material. The first compresses the material to the point $P_A$, with no phase transition, which occurs only in the front of a second shock wave that follows the first. It is easy to show that the propagation velocity of the first shock wave is greater than the propagation velocity of the second shock wave, meaning that there is no superposition between them.[1]

The numerical treatment of dynamic compression of solids, with instantaneous phase transformation, can be made in the following way.[21] An appropriate equation of state is defined for each phase. From point 0 to point A at Figure 1 we use the equation of state of phase 1. Above the point B the equation of state of phase 2 is used. Between points A and B, or between the specific volumes $V_A$ and $V_B$, the pressure is kept constant with value equal to $P_A$.

In the case of plutonium, as the transition from $\delta$ to $\alpha$ phase occurs at very low shock pressures (P>6 kilobars),[14-16] compared to the shock pressure generated by chemical high-explosives (of the order of hundreds of kilobars and, in the process of implosion, in the range of several megabars), a reasonable approach would be simply to put the pressure equal to zero between the specific volumes $V_\delta$ and $V_\alpha$ ($V_\alpha \cong 0.8 V_\delta$), so that the transition would take place instantly without any resistance of material to compression. The advantage of this procedure is that, without significantly changing the results, no equation of state would be required for delta-phase plutonium.

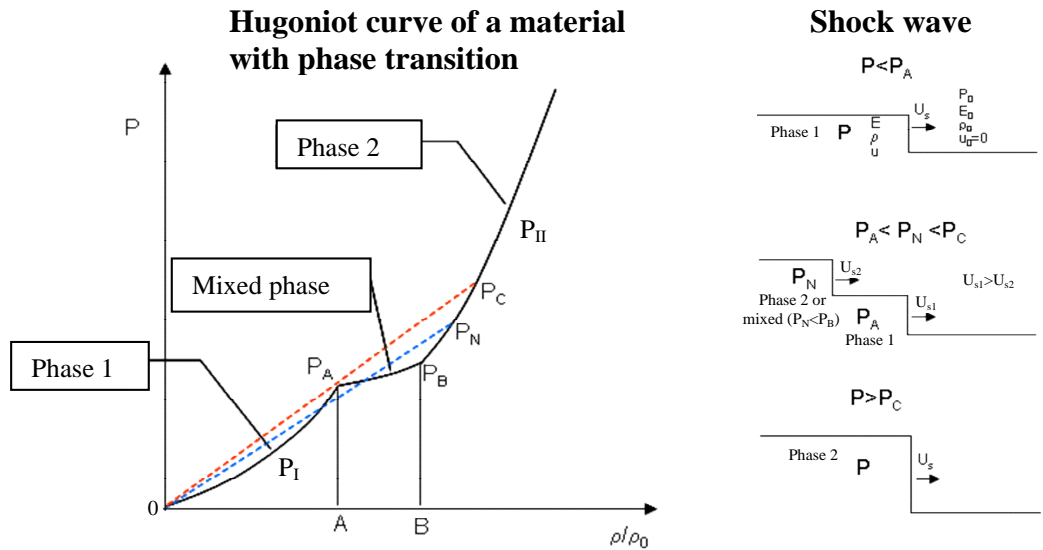

Figure 1: Pressure and density curves in the shock wave in the material with phase transformation. The transition starts from $P>P_A$ and is complete in $P_B$; at the right, we represent the shock waves generated in each case.

The situation is very similar to the already mentioned compression of plutonium in a porous state with initial density equal to the density of plutonium in delta-phase. (In strong dynamic compression of porous materials, the relatively small mechanical strength of the material can be neglected during the closing of the pores.) Of course, just as in the dynamic compression of porous materials, compression of plutonium from its delta-phase produces densities below those that would be produced if plutonium was compressed from its alpha-phase, but still well above those that would be achieved if there were no transformation phase. In fact, as $V_\alpha \cong 0.8 V_\delta$, the densities are not as far away from each other, as will be seen in section 7.

## 4. Properties of actinides. Plutonium allotropic phases

In the periodic table, the main actinides are the elements ranging from the actinium to californium. An important feature of actinides, responsible for many of its properties, is the situation of the electrons in the 5f electronic layer: whether they are localized or itinerant. From protactinium to neptunium, there are electrons in this layer that are itinerant, i.e., they take part in metallic bonds (in the conduction band), while from the americium to californium they are localized, i.e., attached to the atom they belong, therefore not participating in metallic bonds (Fig. 2).

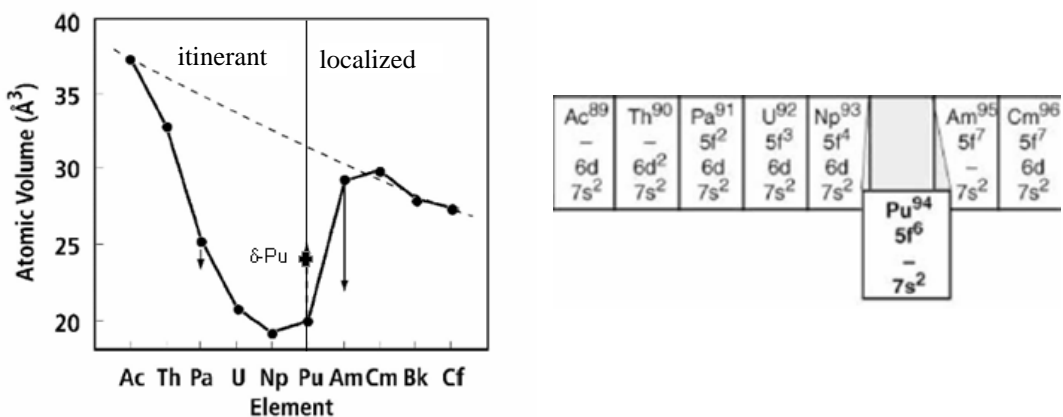

Figure 2: Atomic volume of actinides and characterization of those that have itinerant electrons in 5f shell (so participating of the metallic bonds) and those that have these electrons localized (bonded to the atom they belong). The position of Pu at the border gives it much of its exotic properties.

Plutonium is located just on the border of the elements with these characteristics, a fact responsible for its complex behavior and for many of its exotic properties. The question of the status of 5f electrons in the plutonium atomic structure and its influence on the complex behavior of this element has been very controversial, and only recently, with more accurate quantum studies and through complex computer simulations, it was possible to come to more definitive conclusions about the subject.[22-24]

Figure 3 shows the six allotropic phases of plutonium at normal pressure with their respective densities and transition temperatures, as well as the linear expansion with the temperature in each phase. In the alpha-phase, stable at room temperature, its crystal structure is monoclinic, while in the delta-phase it is face-centered cubic. To give just two examples of the unique properties of plutonium, it should be noted that the density of plutonium in the liquid phase is greater than its density at solidification (behavior similar to the water) and that in the delta-phase the plutonium undergoes slight contraction instead of expansion with temperature increase.

Figure 4 shows the phase diagram of plutonium as a function of atomic percent of gallium added to plutonium.[25,22] As can be seen, from a fraction of about 1.8 at% of gallium, plutonium becomes stable in delta-phase at room temperature (actually theoretically metastable, but with transition so imperceptibly slow that, for all practical purposes, we can consider it as stable over time).

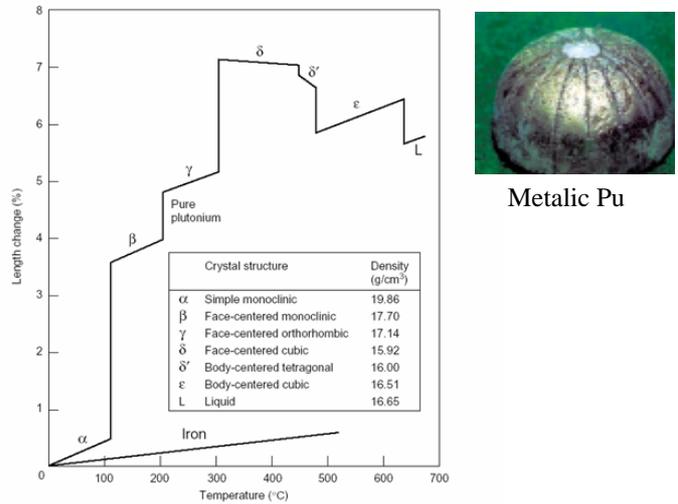
Metalic Pu

Figure 3: Allotropic phases of plutonium and linear expansion versus temperature. Comparison with a common element, iron.

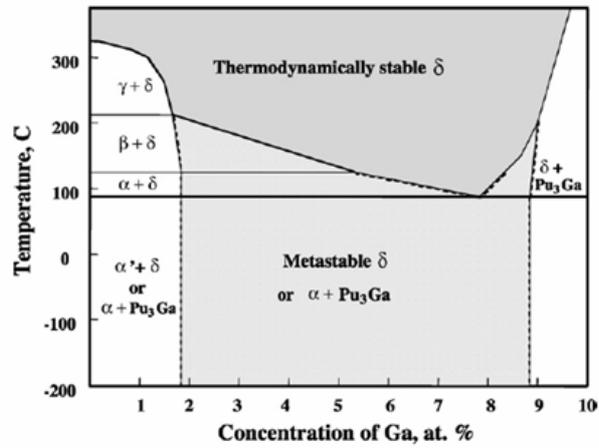

Figure 4: Phase diagram of plutonium versus atomic concentration of gallium.[25]

## 5. Phase change of plutonium under compression

Besides the different allotropic phases of plutonium at normal pressure, it also suffers, like many other elements (it was cited the case of iron), phase transitions when submitted to compression. In many cases, these transitions are somehow associated with variations in the conditions of electrons in the electronic shells as a result of compression. Two illustrative examples of phase transitions with strong and sudden changes in material density are shown in Figure 5 for the actinides americium and curium in function of the static pressure applied.[26,27] These transitions, of course, are associated with "delocalization" of electrons in the 5f shell with the compression, which pass from localized to itinerants, thus participating in the metallic bonds. In the case of uranium, apparently its crystal structure (orthorhombic) remains stable in the pressure range examined (up to 1 Mbar).[28] Some transitions, as can be seen in the figure, are barely perceptible, having little influence on the equation of state (as it is also the case of transition to the liquid phase), while others are much more visible, with large variations in $V/V_0$.

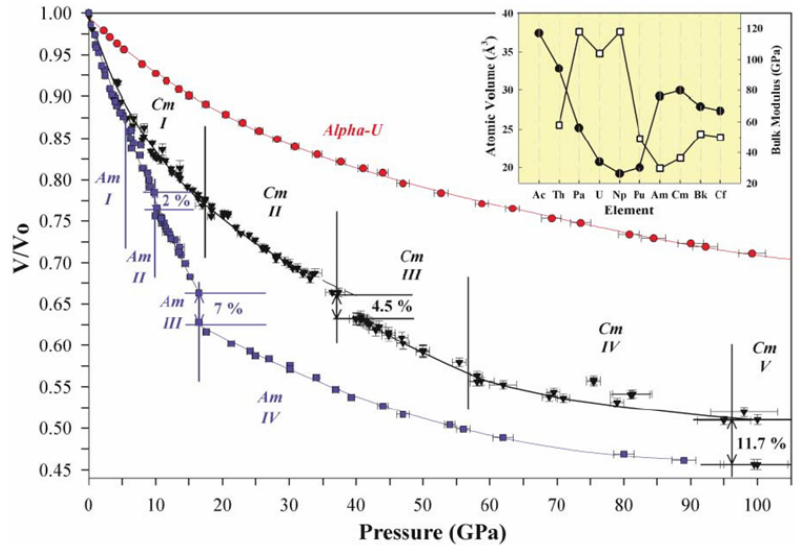

Figure 5: Increase in density of uranium, americium and curium (plotted in terms of specific volume ratio, $V/V_0$) as a function of the static pressure applied.[26] The sudden variations in density are due to changes in the crystal structure of the elements. (See the cited reference for the structures corresponding to different phases.) 1 GPa=10 kbar.

Refined quantum computations (performed on supercomputers) employing the so-called "Density Functional Theory" (DFT)[23] were performed to determine the crystalline structure of actinides (those corresponding to lowest energy) in function of the specific volume (or atomic volume ), which, in turn, is a function of the external pressure applied.[23,24] The results for uranium are shown in Figures 6, and for plutonium in Figures 7 and 8. The calculated energy takes as reference the energy of the bcc (body-centered cubic) crystalline structure, meaning that when the line corresponding to crystalline structure of lower energy crosses the axis at (E-$E_{bcc}$)=0, the crystalline structure acquired is the bcc. For uranium, the crystalline structure remains orthorhombic (lower energy line) until $V/V_0 \cong 0.71$ (P$\cong$1.0 Mbar, see Fig.10), then transiting to the α-Pa structure (alpha-protactinium) and, for higher pressures, to the bcc structure.

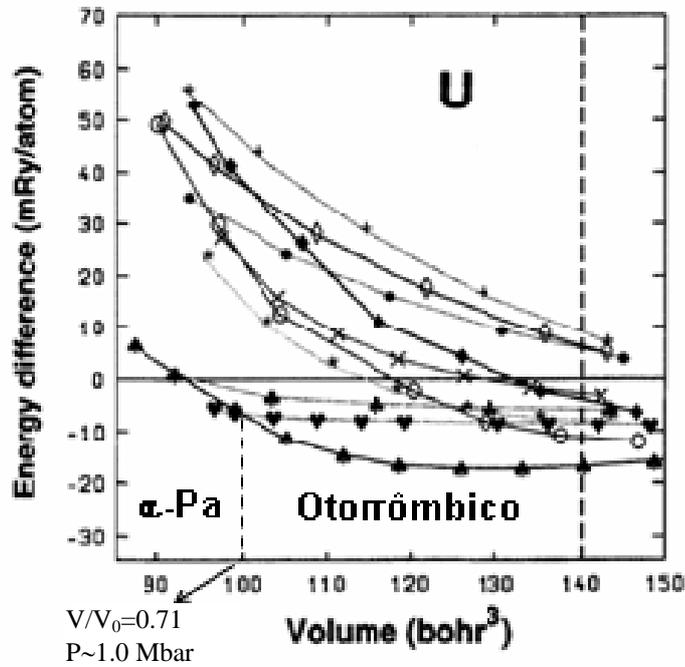

**Figure 6**: Differences in energy for various crystalline structures set for U, taking as reference the energy of the bcc (body-centered cubic) structure, as a function of volume.[24] The vertical dashed line on the right marks the volume corresponding to the normal density of uranium. Down to $V \cong 100$ (bohr$^3$) — a value corresponding to $V/V_0=0.71$ ($P \cong 1$ Mbar, Fig.10) — the lowest energy structure is the normal U orthorhombic structure. Below this value, the structure transits to α-Pa and, for higher pressures, for the bcc. The other crystalline structures, whose energies lie above the energy of the structures here identified, are specified in ref.[24].

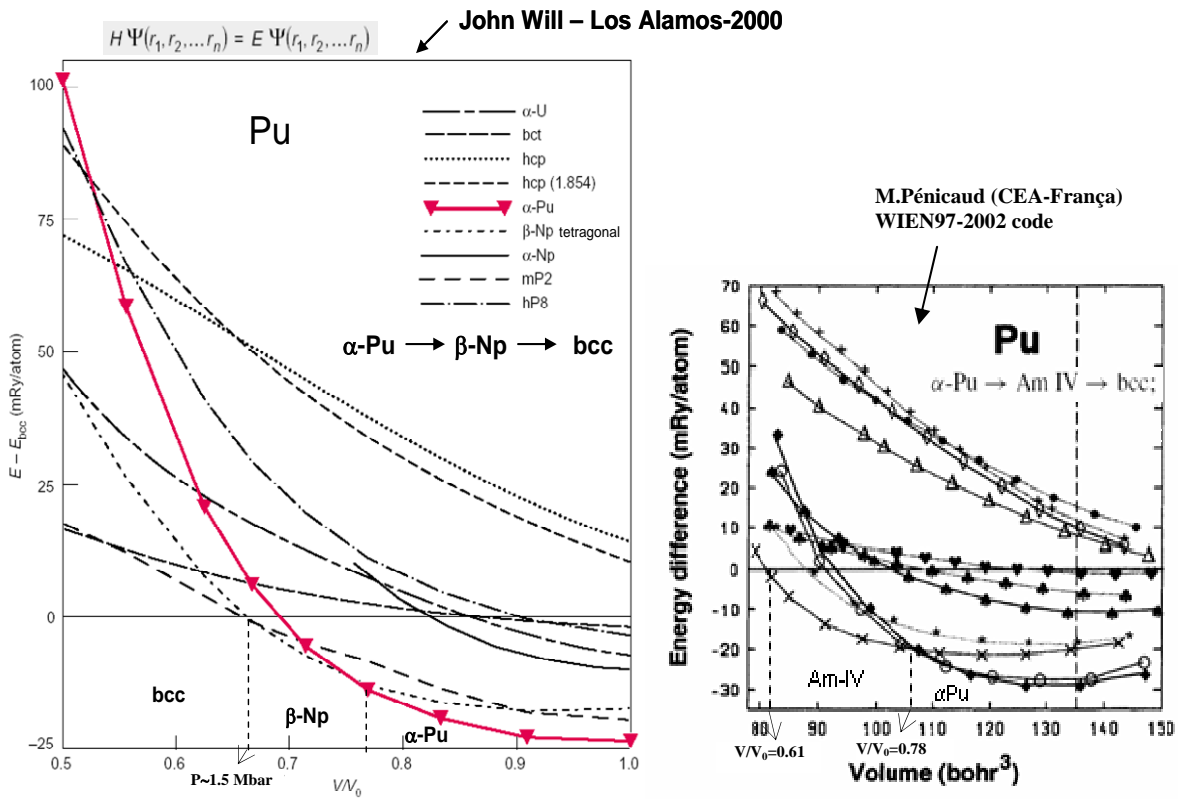

**Figures 7 and 8**: Graphics similar to that of Fig 6 for Pu, obtained by John Will[23] and Pénicaud.[24] The admitted crystalline structures can be seen in the cited references. According to these results, the transitions undergone by Pu (from its alpha-phase) under compression are the following: John Will (α-Pu→β-Np→bcc) and Pénicaud (α-Pu→AmIV(primitive orthorhombic)→bcc). The main differences in the results are in the middle transition and at the point where the transition to the bcc structure occurs. The points where the transitions occur (in terms of specific volume ratio, $V/V_0$) are shown in the figures.

In the case of plutonium, two results with different transitions are presented. Those of John Will (Fig.7), with the following transitions: α-Pu→β-Np→bcc, and those of Pénicaud (Fig.8), with the transitions: α-Pu→AmIV(primitive orthorhombic)→bcc. The points at which the transitions occur are indicated in the figures. The point of the first transition, in $V/V_0 \cong 0.78$, is consistent with experimental results obtained by Dabos et al. (Fig.9),[29] although the hexagonal structure determined by Dabos be questioned and attributed to a probable experimental error.[30]

Despite the difference between the results, the important point to be noted here is that, under high static pressures (in the range of megabars, in which we are most interested) both the uranium and plutonium acquire the body-centered cubic (bcc) structure, being legitimate to admit that in this phase (including, of course, the liquid phase, in strong shock waves) their behaviors are similar, stressing that both have atomic density and volume nearly equal. This means that their equations of state should also be similar, both with respect to the cold pressure as with to the thermal pressure of atoms and electrons.

Thus, it seems legitimate that the equation state of uranium, defined here on the basis of experimental results, be extrapolated to the plutonium, especially in higher pressure ranges of interest (in the order of megabars), typical from those occurring in materials subjected to strong cylindrical or spherical implosion.

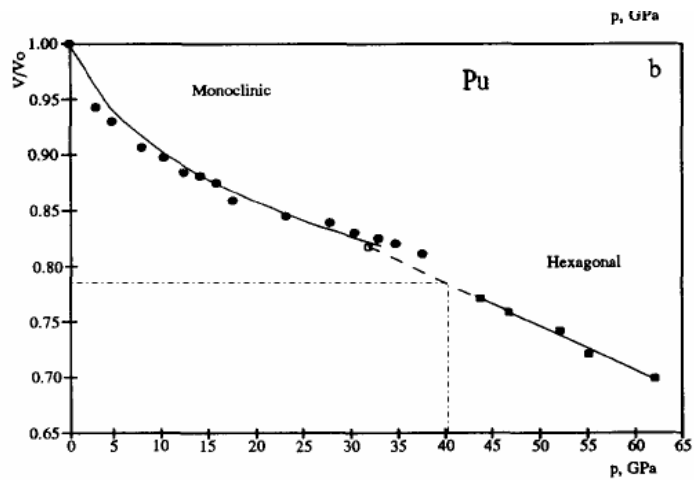

**Figure 9**: Experimental data on static compression of plutonium obtained by Dabos et al.[29] The point where the phase transition occurs coincides with the results in Figs.7 and 8. However, the hexagonal phase has probably been erroneously determined by Dabos et al. (Sika provides a possible explanation of this error.[30])

**6. Data compression of uranium. Equation of state of uranium**

As noted before, data on both static and dynamic compression of uranium are available in the open scientific literature. Figure 10 presents curves of elastic (cold) compression of uranium, $P_c$, obtained by McQueen[4] and Johansson,[5] and Figure 11, Hugoniot curves plotted as a function of points obtained by: 1) Funtikov,[31] Altshuler[32] and Ragan[33] (the latter two, the 2 top points); 2) Johansson (theoretical results, in part);[5] and 3) SESAME library, from Los Alamos.[8,9]

Although the results of McQueen and Johansson in Figure 10 are in close coincidence for pressures below 0.5 Mbar, the data differ significantly for pressures greater than this value. In this work, we opted for the results of McQueen, considering that they allowed a better reproduction of uranium Hugoniot curve plotted as a function of data from Funtikov, Altshuler and Ragan (Fig.11). The Q and q values in the Altshuler expression (11) that best fit the data of McQueen are shown at the lower right corner of Figure 10. Interestingly, the values of the Bulk modulus, $B_0 = \partial P_c / \partial \delta |_{\delta=1} = Q(q-2)/3 = 1.2$ Mbar, its derivative, $B'_0 = (q^2 + 3q-12)/[3(q-2)] = 5.6$ and $\gamma_{g0} = (q^2-6)/[6(q-2)] = 2.3$ are consistent with the experimental values of these quantities for uranium.

In this work we took as reference the data from Funtikov, Altshuler and Ragan (Fig.11), which allow a plot of Hugoniot curve up to extremely high pressures. The two top points were obtained by Altshuler and Ragan in experiments with nuclear explosions, while the lower points (below 12.63 Mbar) were obtained by Funtikov. Note that the values match perfectly, allowing a smooth plotting of the curve. The Johansson-Persson values were calculated using the three-term equation of state, with the pressure of electrons given by eq.(10) (with $\beta_0=176$ erg/gK$^2$) and with extrapolation of the experimental curve $P_c(\rho)$ to bring it closer to the 0 K isothermal curve obtained by the Thomas-Fermi-Kalitin theory for pressures above 100 Mbar.[5] As can be seen in Figure 11, the Hugoniot curves agree closely for pressures up to about 4 Mbar.

The parameters of the equation of state (1) which best fit the data from Funtikov, Altshuler and Ragan are shown at the right of the graph in Figure 11. Q=0.359 Mbar and q=12.028, as already mentioned, were defined according to the data of McQueen (Fig.10); $\gamma_{g0}$=2.2, the Gruneisen coefficient of uranium at the normal density, was obtained from ref. [7] ; $\gamma_e$=0.5 — the controversial electronic Gruneisen coefficient — was put equal to its traditional value, extracted from the Thomas-Fermi theory; $\beta_0$=75

erg/gK$^2$, the initial electronic specific heat coefficient, was defined so that the equation of state (1) (using the expression (10)) reproduces exactly the Funtikov point corresponding to P=12.63 Mbar (the value obtained is very close to the value admitted in ref.[34]); the constant b=4.5x10$^6$ erg/gK was set so that the equation of state reproduces the result of the Thomas-Fermi model for an extreme point obtained from the table in the Appendix 5 of ref.[35]; finally, the constant $\ell = 1.0$ was defined to better reproduce results from the compression of porous uranium (see Fig.12).

As can be seen in Figure 11, there is an almost perfect agreement between the calculated points by LUI1[36] code, using the equation of state defined here for uranium, and the Hugoniot curve plotted using the data from Funtikov, Ragan and Altshuler.

Another result that confers great reliability to the equation of state (1), with the parameters set forth herein, is the almost perfect agreement between the temperature values calculated by this equation of state and the temperature values shown in Figure 12 (obtained from Lomonosov's transparencies[37]), both for shock compression of uranium at solid density (m=1) as for shock compression of uranium with an initial porosity of m=2 ($\rho_0$=9.5 g/cm$^3$). Results were calculated at points where the isothermal curves cross the Hugoniot curves.

Comparison with results for uranium with higher porosities (m> 2), in Figure 12, is difficult due to the high degree of uncertainty in these results.

Figure 13 shows the uranium temperature curve plotted in function of the shock pressure. Similarly, the results were calculated by the LUI1 code with the equation of state defined in this work. The melting curve of uranium shown in Figure 13 was obtained using the known relationship between the Gruneisen coefficient and the Debye temperature $T_D(\rho)$,[35]

$$\gamma_g = \frac{\rho}{T_D}\frac{dT_D}{d\rho}, \qquad (12)$$

and the expression below, from the Lindemann rule,[38]

$$T_m = T_{m0}\frac{T_D^2}{T_{D0}^2}\frac{\rho_0^{2/3}}{\rho^{2/3}}. \qquad (13)$$

Solving eq.(12), with $\gamma_g$ given by (7), we obtain:

$$\frac{T_D}{T_{D0}} = (\frac{\rho}{\rho_0})^{2/3}\exp[(\gamma_{g0} - 2/3)(1 - \rho_0/\rho)]. \qquad (14)$$

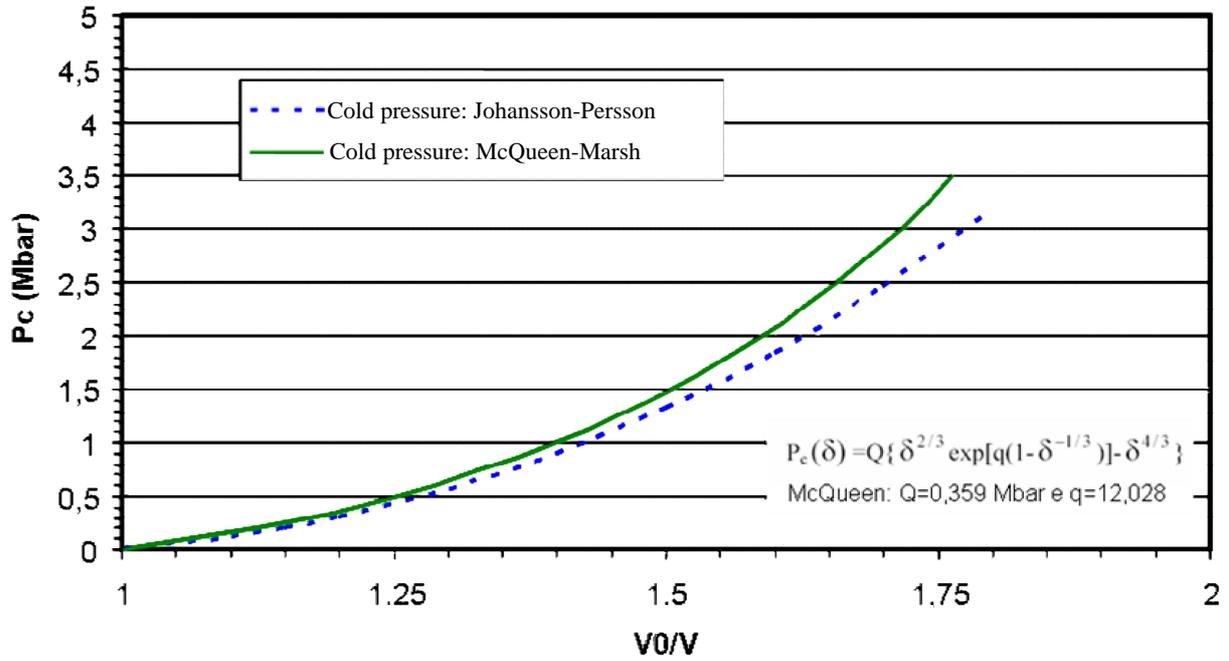

**Figure 10:** Elastic ("cold") pressure curves versus increase in uranium density obtained by McQueen-Marsh[4] and Johansson-Persson.[5] The values of the constants Q and q in the Altshuler expression (11) that best fit the data of McQueen are shown at the bottom right of the graph. These values are used in this study.

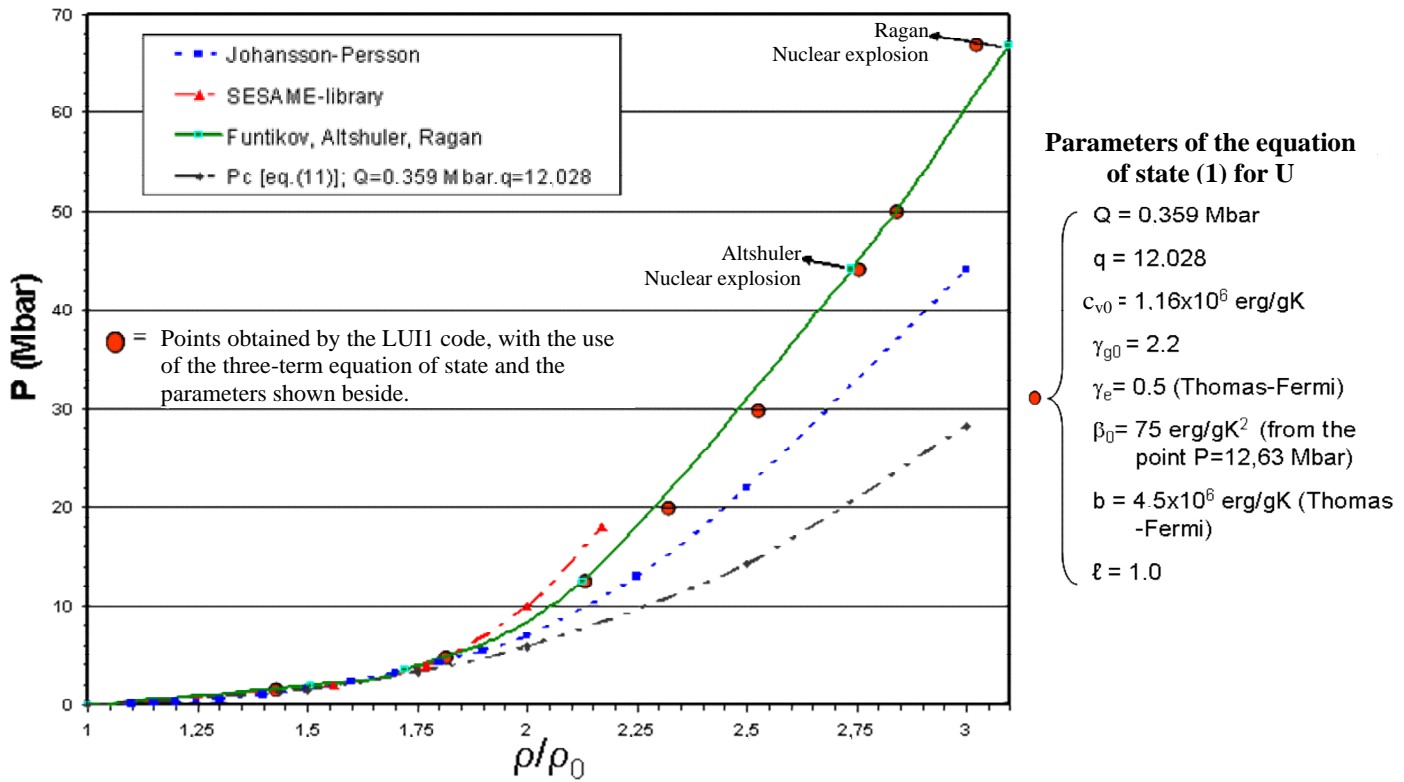

**Figure 11:** Hugoniot curves of uranium based on data obtained by 1) Funtikov,[31] Altshuler[32] and Ragan[33] (the latter two, the 2 top points); 2) Johansson-Persson;[5] and 3) SESAME library, from Los Alamos.[8,9] In the graphic we present also the calculated points in this work, using the LUI1 code[36], the equation of state (1) and the parameters set on the right. Note the excellent agreement with the curve based on data from Funtikov, Altshuler and Ragan, considered the most reliable and taken as reference in this work.

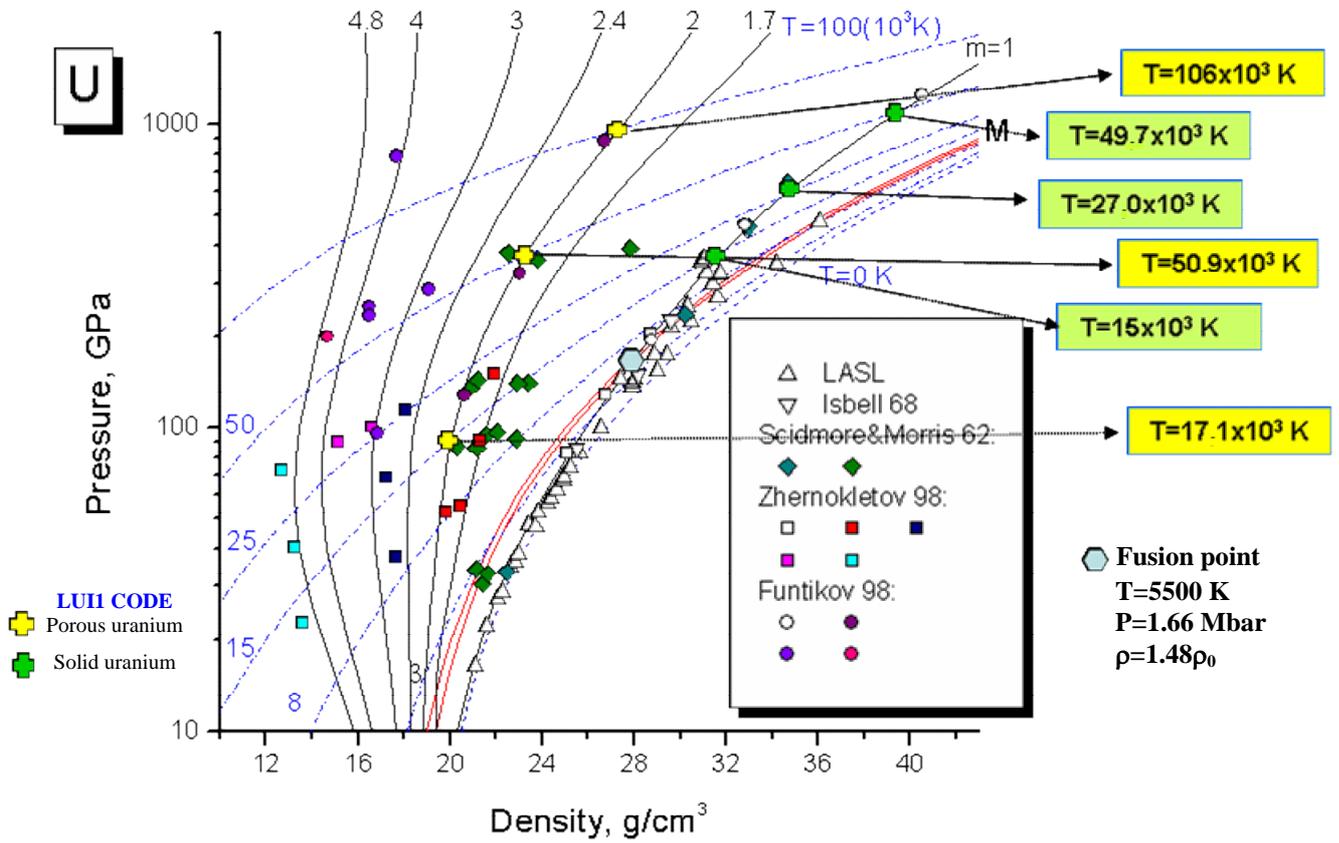

**Figure 12:** Data of dynamic compression of porous uranium obtained by different authors and presented by Igor Lomonosov;[37] $m=V_{00}/V_0$ is the initial porosity. The points where the isotherms (blue) intersect the Hugoniot curves give the temperature value at these points. Points calculated by the LUI1 code, using the equation of state here defined, are presented for m=2 ($\rho_0$=9.5 g/cm$^3$) and m=1. Note the excellent agreement with the temperature values, giving greater reliability to the parameters of the equation of state (1) here determined.

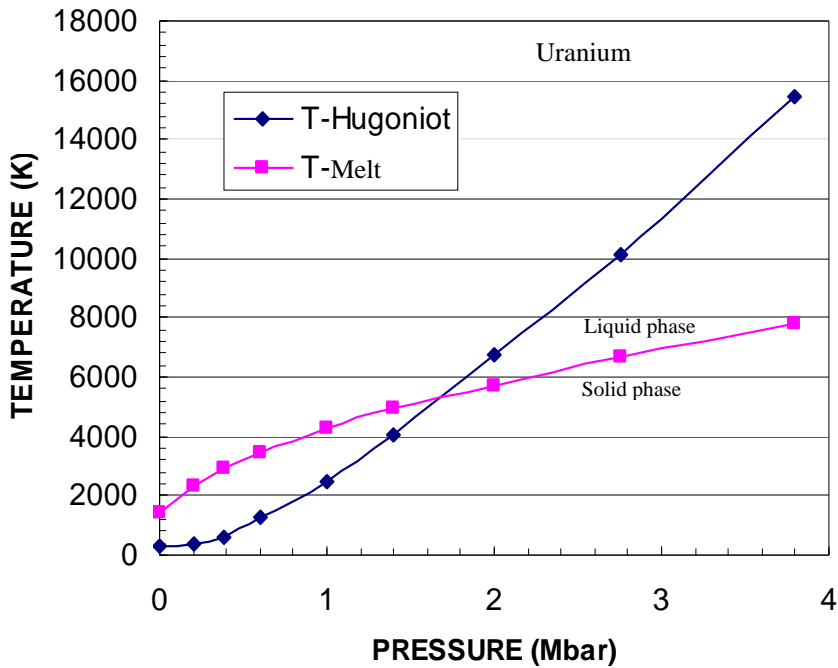

**Figure 13:** Temperature curve for uranium versus shock pressure, calculated by the LUI1code with the equation of state (1). The melting curve (eq.(15)) intersects the Hugoniot curve at P≅1.66 Mbar. Above this pressure, the uranium changes to the liquid phase.

Substituting (14) into (13),

$$T_m = 1.1 T_{m0} \left(\frac{\rho}{\rho_0}\right)^{2/3} \exp[2(\gamma_{g0} - 2/3)(1 - \rho_0/\rho)], \quad (15)$$

where $T_{m0}$ is the melting temperature of the material at standard conditions (1406 K for U and 912.5 K for Pu); the 1.1 empirical factor is a small correction in the melting temperature at the solid density $\rho_0$.[39] As it is known, $T_m$ rises with the ratio $\rho/\rho_0$, which in turn depends on the shock pressure on the material (Fig.11).

Note that the transition to the liquid phase (the melting point) occurs in P≅1.66 Mbar, corresponding to density ratio of $\rho/\rho_0$=1.47 and temperature T=5400 K, values very similar to those reported by Lomonosov[37] (Fig.12).

## 7. Equation of state of plutonium. Comparison with AWE

As shown by the results presented in section 5, plutonium and uranium tend toward the same final solid allotropic phase (bcc) when submitted to high static pressures (in the range of megabars) or to the liquid state in strong shock waves. These facts, along with the fact that both elements have atomic volume and density nearly identical, suggest that their equations of state should have similar behavior in this pressure range, thus justifying the use, in the equation of state of plutonium, of the same parameters defined here for the equation of state of uranium (Fig.11), with modifications only in the value of the specific heat and in the value of the not so much influential constant ℓ.

Unfortunately, unlike the case of uranium, there are no unambiguous data of dynamic compression of plutonium in the open scientific literature, making it difficult to establish with certainty the hypothesis considered above.

Figure 14 shows the compression (Hugoniot) curves of α and δ phase plutonium (the last stabilized at room temperature) calculated by our LUI1 code in the pressure range of interest. The compression of δ phase plutonium occurred by the procedure explained in section 3, i.e., by considering pressure simply equal to zero during the time the plutonium takes to reach its normal α phase density.[(*)] This is justified, as noted in section 3, by the fact that the transition from δ to α phase takes place at very low shock

---

[(*)] Calculations performed using a simple equation of state for the Pu in delta-phase (with data taken from ref. [16]) and a pressure transition of 6 kbar, did not significantly alter the results.

pressures comparing to the pressures here considered. (As mentioned in section 3, the situation is similar to the compression of porous plutonium with density equal to the density of δ phase plutonium.) Thereafter, the pressure is given by the equation of state defined in this work and also assuming that the plutonium quickly changes to the bcc (body-centered cubic) phase and, subsequently (or directly, depending on conditions), to the liquid phase, in the pressure range of interest (Fig. 14).

The first two curves in Figure 13 are plotted taking as reference the density of plutonium in its normal α phase ($\rho_0=1/V_0=19.8$ g/cm$^3$), while the rightmost curve represents the same compression of plutonium in δ phase but taking as reference the nominal density of plutonium in this phase ($\rho_0=15.9$ g/cm$^3$).

Figure 14 makes clear that, for a given shock pressure, the final density of plutonium compressed from its δ phase is lower than the density reached by compression it from its α phase. This is an obvious fact, since in the compression of a porous material (as well as in the compression of materials with abrupt phase change), there is a greater increase in entropy and temperature. Hence, in the Pxρ diagram, Hugoniot curves of porous materials lie at left of Hugoniot curves of normal solid density materials.[2]

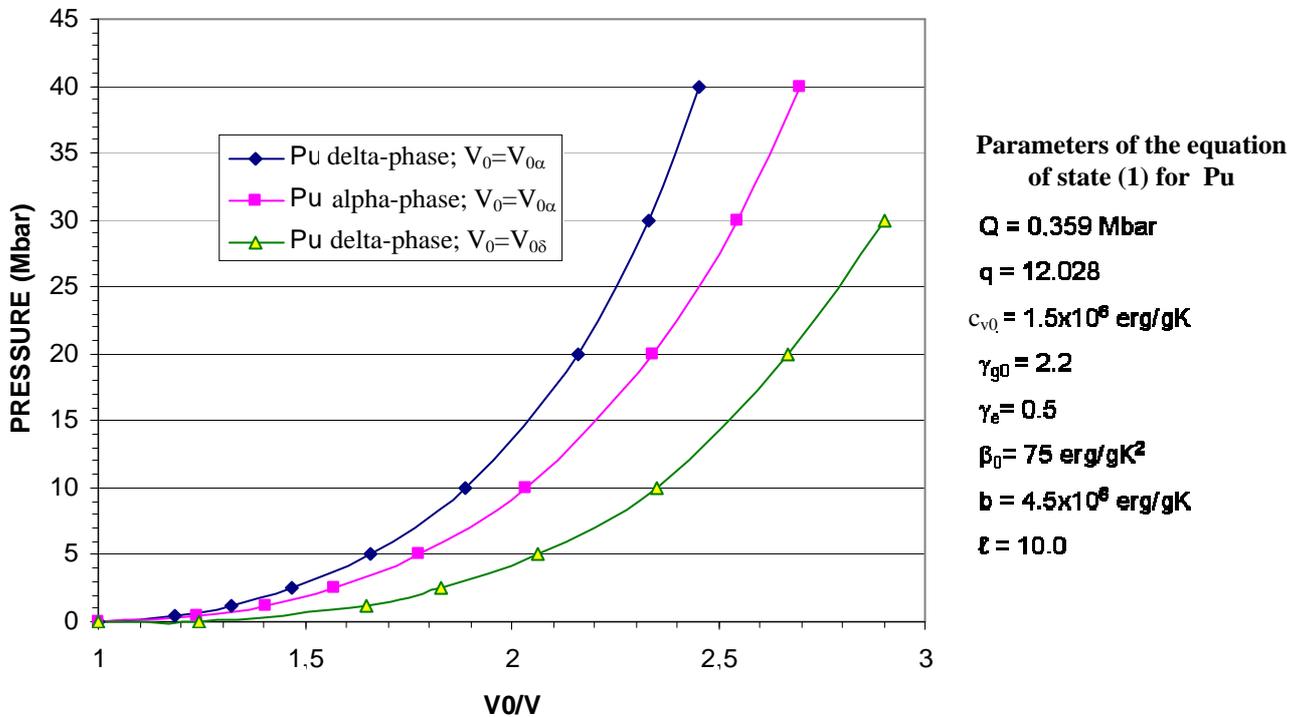

**Figure 14:** Dynamic compression (Hugoniot curves) of α and δ phase plutonium, taking as reference the initial density of plutonium in α phase ($\rho_0=1/V_0=19.8$ g/cm$^3$). The rightmost curve is a plot of the compression of δ-stabilized plutonium taking as reference the nominal specific volume of this phase ($\rho_0=1/V_0=15.9$ g/cm$^3$). The results were obtained by the LUI1 code, using the equation of state (1) with the same parameters specified in Fig.11 for uranium, but with different values of $c_{v0}$ and $\ell$.

Although we have no explicit data on the compression of plutonium at high pressures range, ref.[40] presents data on the compression of a material that we presume is the plutonium, although in the article of the cited reference this is not explicitly stated (it seems that the same perception had Barsamian[41]). The authors of this paper are M. Pollington, P. Thomson and J. Maw, from the English laboratory AWE (Atomic Weapons Establishment). The data presented by them are shown in Figure 15, including what is supposed to be the Hugoniot curve of plutonium compressed from its δ phase, as demonstrated by the excellent agreement with the compression data of plutonium calculated by the equation of state here defined (Fig.14). Similarly as in Figure 14, the Hugoniot curve of Pollington et al. is plotted as a function of $V/V_0$, where $V_0$ is believed to be the specific volume of plutonium in α phase. (It seems that the other curves are plotted taking $V/V_{0\delta}$ as reference.) The reason to assign a value of $\ell = 10$ to the constant in the expression (8) was due to the fact that this value allowed a better agreement at the extreme of the curve, for pressures above 20 Mbar (note that the melting temperature of Pu is much smaller than that of U).

Figure 16 shows the Hugoniot temperature curve of plutonium compressed from its α and δ phases as a function of the applied shock pressure, as well as the transition curve for the liquid phase (eq.(15) and Fig.14). Note that in the case of α phase, comparing with the results in Figure 7, the plutonium change to the liquid phase before transiting to the bcc phase, and in the case of δ phase, due to much higher temperature reached, even before transiting to β-Np (or Am-IV) phase.

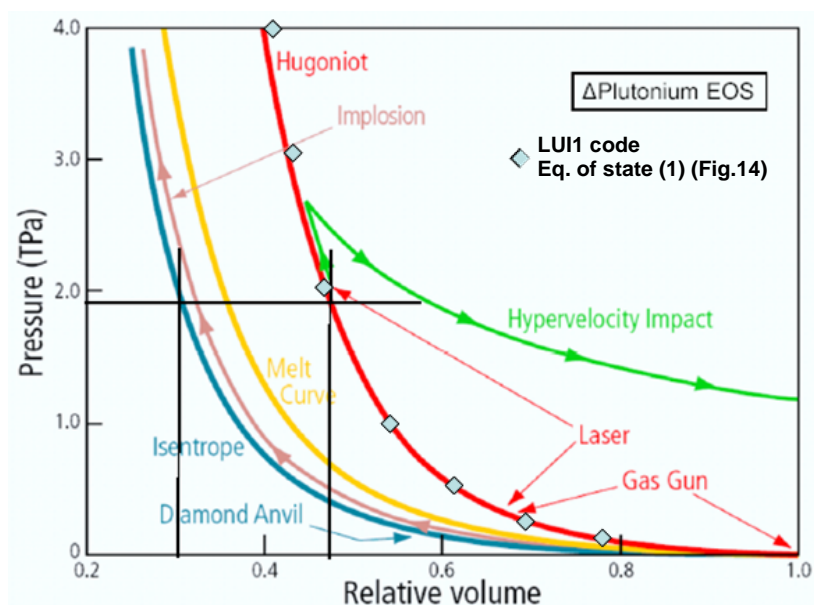

**Figure 15:** Comparison with results presented by Pollington et al.,[40] supposedly for the plutonium. (The information that this is the compression of plutonium in its δ phase comes from Barsamian.[41]) The analysis of the results leads to the assumption that the Hugoniot curve is plotted in terms of $V/V_{0\alpha}$, while the other curves are in function of $V/V_{0\delta}$. 1 TPa = 10 Mbar.

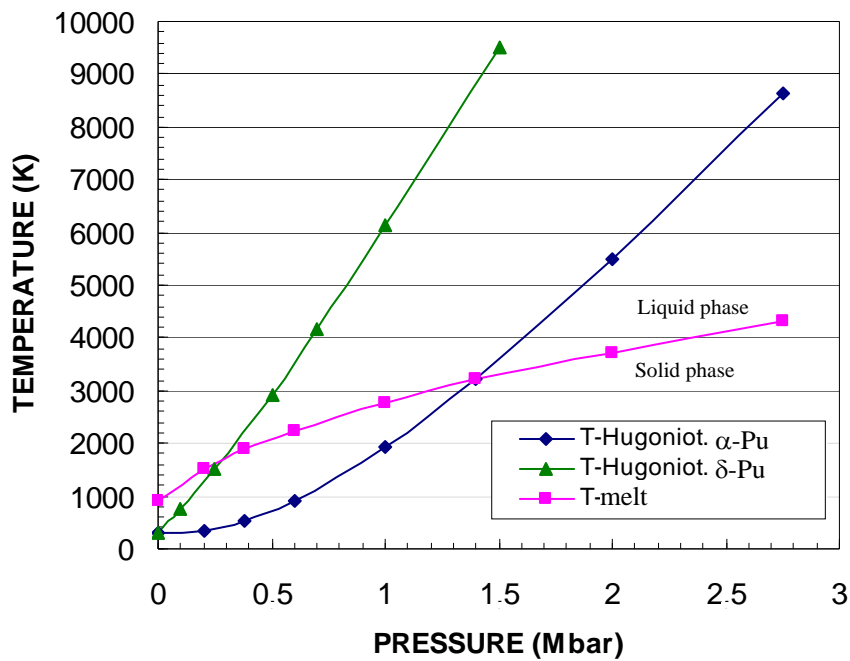

**Figura 16:** Hugoniot curves for temperature of plutonium ($\alpha$ and $\delta$ phases) as a function of the applied shock pressure, calculated by the LUI1 code with the equation of state (1). The melting curve (eq.(15)) intersects the Hugoniot curves at P$\cong$1.40 Mbar and T $\cong$3200 K in the case of $\alpha$ phase, and at P$\cong$0.28 Mbar and T$\cong$1700 K in the case of $\delta$ phase. Above these pressures, in the respective cases, the plutonium changes to the liquid phase.

## 8. Simulation of the "Fat Man" atomic bomb using the Pu equation of state

Reports come from secret files released after the collapse of the Soviet Union, as well as research conducted by John Coster-Mullen,[42] in the United States, allow us to assume as true the configuration of the atomic bomb "Fat Man" (detonated in Trinity and Nagasaki) presented in Figure 17 (see http://nuclearweaponarchive.org).

We do not discuss here in detail the reasons for this configuration, but, briefly, the purpose of the fissile mass (Pu) is obviously to become supercritical and cause a nuclear explosion; of the tamper of natural uranium is to provide inertia and to reflect neutrons back into the fissile mass; of the borated acrylic (the $B^{10}$ is highly neutron absorber) is to absorb neutrons moderated and reflected from the aluminum and the high explosive, improving the conditions for the initial subcritical configuration and decreasing the probability of pre-ignition; and, finally, of the aluminum is to minimize the effect of the rarefaction wave coming from the pressure drop in the high explosive after the impact of the detonation wave, to increase the effect of shock wave convergence and also to minimize the effects of the Rayleigh-Taylor instabilities in the explosive-solid interface.

This configuration will be used to test the equation of state (1) defined here for the plutonium, through the use of two coupled computer codes developed by the author of this work: the first already aforementioned, LUI1 code, simulates the implosion of the fissile mass by the spherical detonation of the high-explosive surrounding the system. It

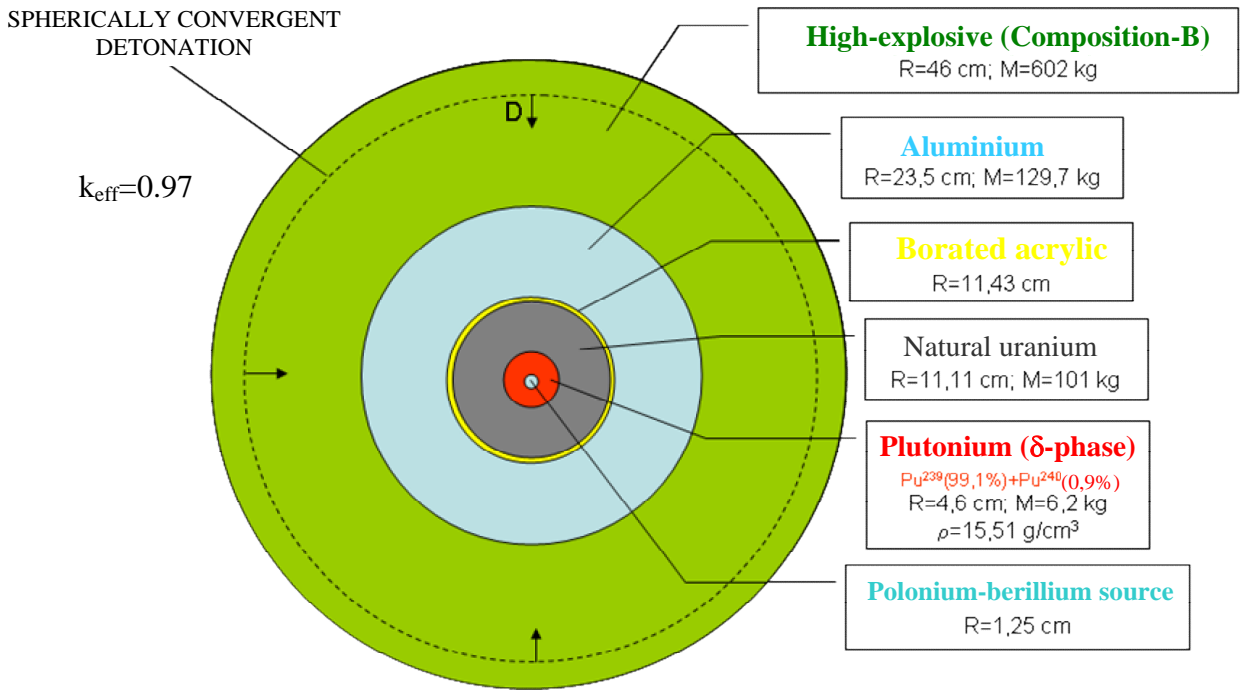

**Figure 17**: Possible configuration of the atomic bomb "Fat Man", based on data reported by Carey Sublette (see, on the internet, http://nuclearweaponarchive.org). Not shown in the figure are the explosive lenses responsible for generating the spherically convergent detonation wave into Composition B, since they have no significant influence on the process of internal fissile mass implosion. The criticality of the initial configuration ($k_{eff}$=0.97) was calculated by the LUI1 code.

Parameters of the equation of state (1) for Be and Al, based on data obtained from refs.[4,43] and [2], respectively. For the acrylic (whose influence might have been neglected) it is used the Gruneisen equation of state and the data from refs.[4,8]; and, for composition B, it is used the gamma-law equation of state.[3,44]

**Beryllium**
$$\begin{cases} Q = 0.917 \text{ Mbar} \\ q = 5.902 \\ c_v = 1.75 \times 10^7 \text{ erg/gK} \\ \gamma_{g0} = 1.16 \\ \gamma_e = 0.5 \\ \beta_0 = 190 \text{ erg/gK}^2 \\ b = 0.0 \quad \text{(when b=0, the eq.(10) is used)} \\ l = 0.0 \end{cases}$$

**Aluminium**
$$\begin{cases} P_c = \text{a 5-degree polynomial adjusted to experimental results} \\ c_v = 9.234 \times 10^6 \text{ erg/gK} \\ \gamma_{g0} = 2.025 \\ \gamma_e = 0.5 \\ \beta_0 = 518 \text{ erg/gK}^2 \\ b = 0.0 \\ l = 0.0 \end{cases}$$

**Acrilic**: Gruneisen equation of state: $P = P_H + \gamma_g \rho(E - E_H)$, $\gamma_{g0} = 1.0$, $U_s = 2.58 \times 10^5 + 1.42U$

**Composition B**: Gamma-law equation of state: $P = (\gamma-1)\rho E$, $\gamma = 2.77$ (D=8 km/s), $\rho_0 = 1.7$ g/cm³

calculates also the criticality ($k_{eff}$) and generates data on the system configuration in successive hydrodynamic cycles specified by the user during the implosion. For the plutonium and uranium tamper the equations of state defined here will be used, while for the other elements the equations of state are defined on page of Figure 17.

The second code, RIC1,[45] performs the calculations of the fissile mass explosion at certain criticality points or configurations specified by the user during the implosion. The code solves numerically the time-dependent neutron transport equation coupled to radiation-hydrodynamics equations. Its present version has an important option to solve the transport equation (with an additional term due to a moving system in lagrangian coordinates[46]) by the characteristic method[47,48] to overcome the negative-flux problem.

Both codes make use of the Hansen-Roach cross sections[49], whose precision, for the system in focus, may be proved in the benchmarks shown in Table 1. This table presents results of criticality calculations performed on fast neutron systems (some experimentals[50]), using the codes here employed and also, for comparison, the KENO-IV[51] and ANISN[52] codes.

The results of the "Fat Man" simulation, obtained by using the LUI1 and RIC1 coupled codes, are shown in Figure 18, where we seek to summarize, in one graph, the most relevant informations.

First, on the graphic, we show the path of the convergent detonation wave in the Composition B and the path of the implosion shock wave (generated by the impact of the detonation wave on aluminum) in the various material layers, until the time the shock takes to reach the center of beryllium, where it is reflected. The density, pressure and (negative) implosion velocity of the fluid are shown at some relevant points.

Next, it is shown the criticality insertion, ranging from the initial value of $k_{eff}$=0.97 to the maximum value of $k_{eff} \cong 1.51$. Note that $k_{eff}$ varies significantly only from the moment the shock wave hits and compresses the natural uranium, and, more markedly, after its impact on the plutonium, whose compression is more heavily influenced by the cumulative hydrodynamic effects characteristic of a convergent shock wave.

In Figure 18 we show also the energy yield of the nuclear explosion calculated to several configurations corresponding to different points of criticality, and the region of time or the region of maximum criticality reached where nuclear explosion in the "Fat Man" must have occurred. (It should be noted, however, that due to the time the multiplication of neutrons takes to generate destructive energies into the fissile mass, a persistent fission chain reaction (ignition) probably occurred a short time before, as

**Table1:** Criticality calculations in fast neutron systems (cases 3, 4 and 6 from ref. [50]) performed by KENO-IV,[51] ANISN[52] and LUI1[36] (RIC1[45]) codes, using the Hansen-Roach cross sections.[49]

**HANSEN-ROACH CROSS SECTIONS – 16 GROUPS, $P_0$-$\sigma_{tr}$**

| TESTS | $k_{eff}$ KENO-IV | ANISN(*) | LUI1-RIC1(*) | |
|---|---|---|---|---|
| 1 | 1.314 | 1.317 | 1.312 | (*) $S_8$ |
| 2 | 1.0001 | 1.0052 | 1.0035 | |
| 3 | 1.0037 | 1.0007 | 0.9955 | 1.00298 |
| 4 | 0.9801 | 0.9808 | 0.9791 | Vellozo: SCALE, 238 groups[53] |
| 5 | 2.0464 | | 2.0390 | |
| 6 | | | 0.9983 | |

1 - $U^{235}$(93%;R=10.5 cm;18.81 g/cm$^3$) + 5 cm of U-nat

2 – $U^{235}$(93%;R=6.936 cm;18.81 g/cm$^3$) + 4.75 cm of U-nat (critical system)

3 - Flattop-25 ($U^{235}$(93.3%);R=6.116 cm) + 18.01 cm of U-nat (critical)

4 - Big Ten: [$U^{235}$(10.17%)+$U^{238}$(89.72%)+$U^{234}$(0.105%);R=30.48 cm] + 15.24 cm of U-nat

5 - Idem Big Ten $U^{235}$(100%)

6 - Jezebel [spheric Pu ($\delta$-phase);R=6.385 cm;15.6 g/cm$^3$]

**IMPLOSION SIMULATION OF NAGASAKI BOMB ("Fat Man")**

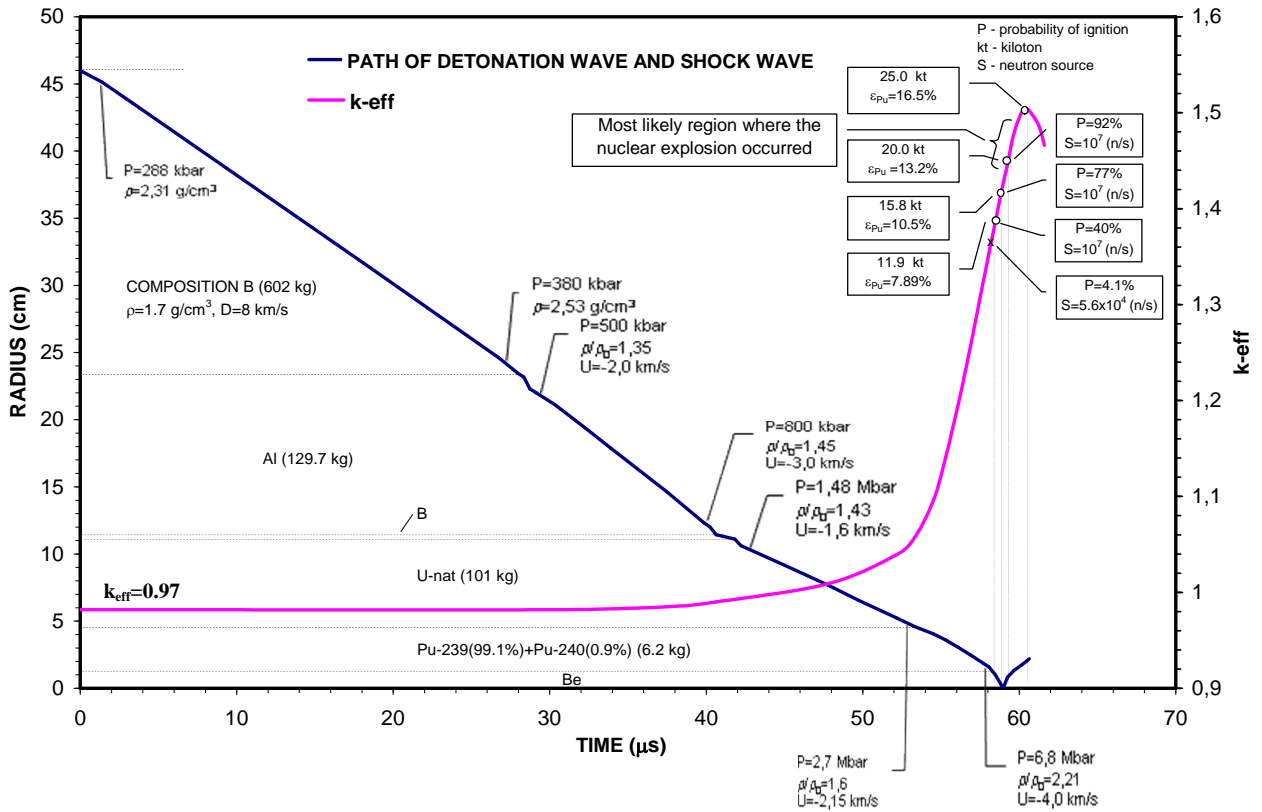

**Figure 18:** Results of numerical simulation of the atomic bomb "Fat Man" performed by using the LUI1 and RIC1 coupled codes. The calculation of nuclear explosion is made in several configurations corresponding to different points of criticality. 1 kt=1 kiloton; $\varepsilon_{Pu}$ is the efficiency of the explosion in terms of Pu burning (21% of the energy of the explosion is produced by fast fissions in natural-U tamper).

indeed is indicated by the ignition probabilities presented in Fig.18.) This region is consistent with the estimated range of energy released by the "Fat Man" explosion, situated between 18 and 24 kilotons.[54] As can be seen in the figure, this time interval is situated just ahead the instant the shock wave reaches the inner artificial neutron source, represented by the beryllium (actually, a source type polonium-beryllium), whose purpose is to produce a neutron pulse (with a estimated valued of $10^7$ n/s) responsible for generating a persistent chain reaction in the fissile mass.

The ignition probability values were calculated by the formula:[3]

$$P(t) = 1 - \exp[-2S/(\overline{\upsilon}\Gamma)\int_0^t \rho(t')dt'], \qquad (16)$$

where S is the neutron source (n/s), $\overline{\upsilon}^{Pu} = 3$ is the average number of neutrons released per fission, $\Gamma \cong 0.8$ is the Diven factor and $\rho(t) = [k_{eff}(t) - 1] / k_{eff}(t)$ is the reactivity.

Note that, during the implosion, before the shock wave reaches the inner neutron source (causing the presumed mixture of polonium with beryllium and the ($\alpha$, n) reactions), the source taken was $S^{240} = 5.6 \times 10^4$ n/s, a value corresponding to random neutron source formed by the spontaneous fissions of $Pu^{240}$ presented in the plutonium mass used (6.2 kg) and whose isotopic composition is specified in Figure 18. With this value of S, the probability of an undesirable persistent chain reaction (ignition) until the instant marked with an "x" on the graphic was calculated to be only 4.1%.

With the activation of the internal neutron source, after being impacted by spherically converging shock wave, we adopted the value $S = 10^7$ n/s characteristic of a typical polonium-beryllium neutron source with the specified dimensions. The ignition probability at various moments following gives the chances of occurring a persistent fission chain reaction into the fissile mass after activation of this neutron source (Fig. 18).

Figures 19 and 20 show the spatial distribution of pressure, density and temperature in the natural uranium and plutonium, respectively soon after the collision of the shock wave onto the internal beryllium neutron source (at t=58.88 μs) and soon after its reflection at the center (at t=59.31 μs). Note that the plutonium is compressed from its δ phase and there is a strong density variation due to the sudden change (with no resistance) to the α phase.

The contributions of the elastic (cold) pressure and of the thermal pressures of atoms and electrons to the total pressure in the plutonium, at time t=58.88 μs, can be seen in Figure 21.

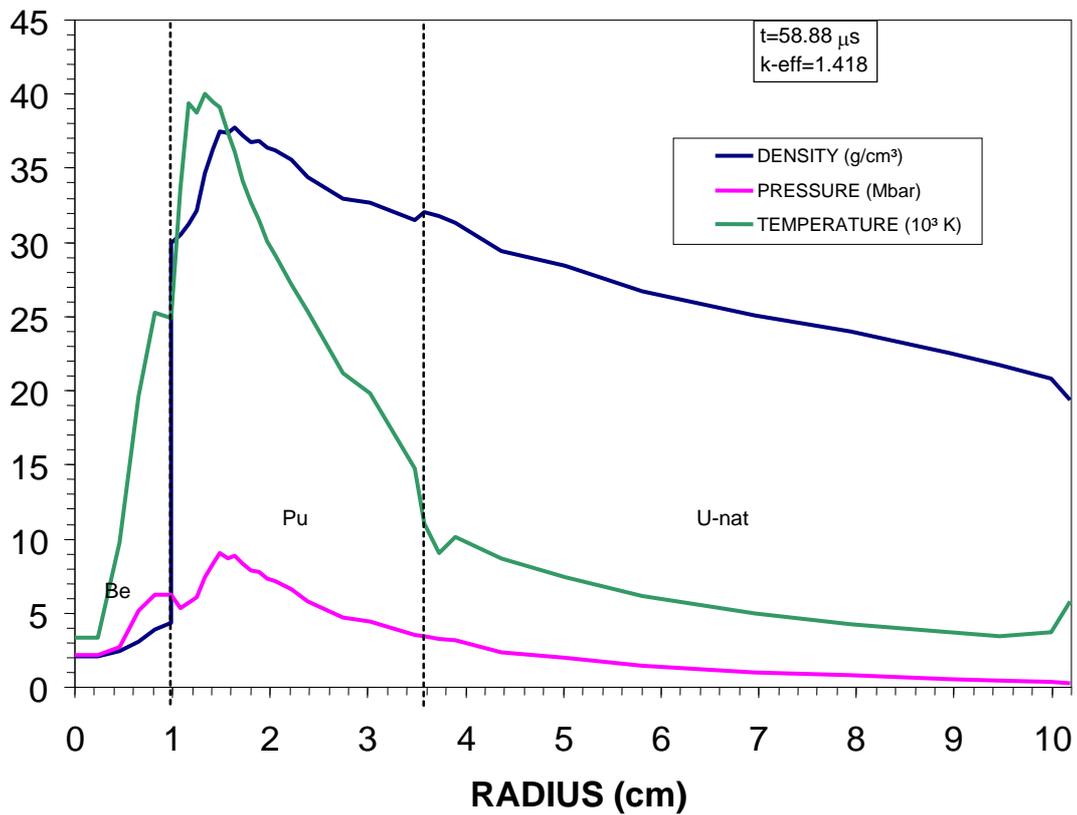

**Figure 19:** Spatial distribution of pressure, density and temperature in the Pu and natural-U, at t=58.88 μs, shortly after the impact of the spherically convergent shock wave on the beryllium. The Pu is compressed from its delta-phase. The nuclear explosion with releasing energy of 15.8 kilotons (Fig.18) was calculated in this configuration.

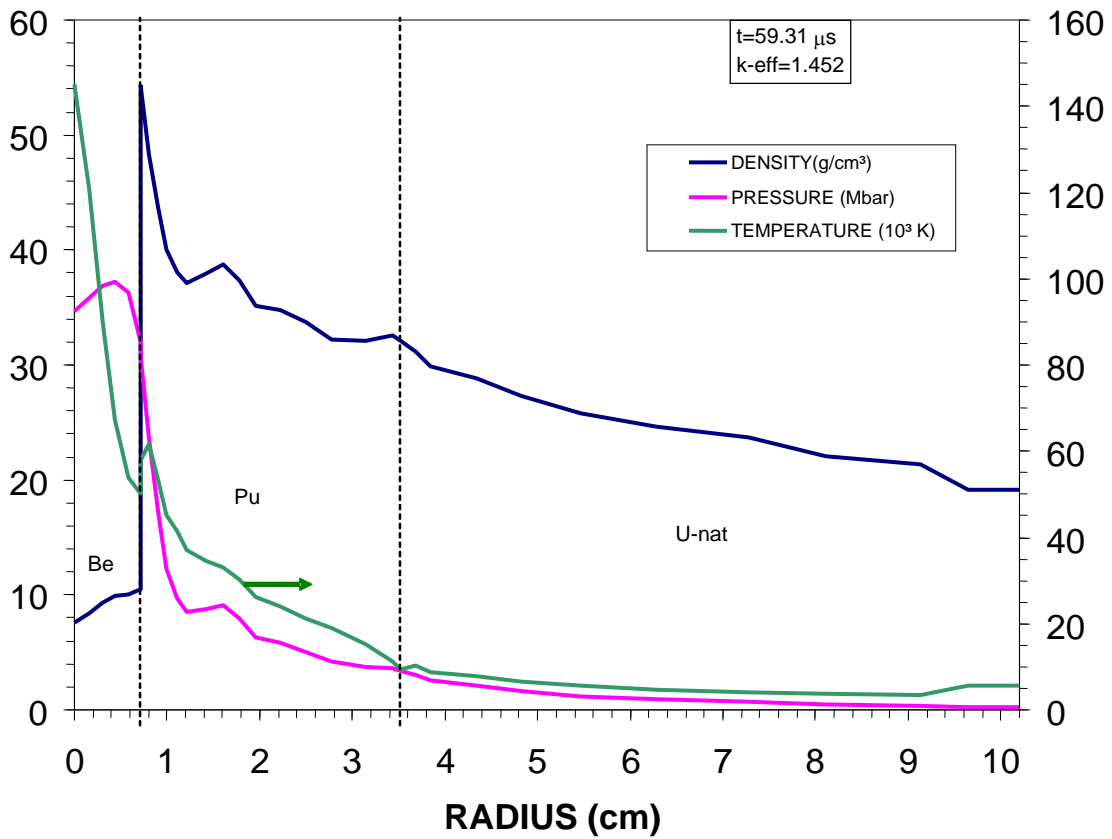

**Figure 20:** Spatial distribution of pressure, density and temperature in the Pu and natural-U, at t=59.31 μs, after reflection of the shock wave at the center. The nuclear explosion with releasing energy of 20 kilotons (Fig.18) was calculated in this configuration.

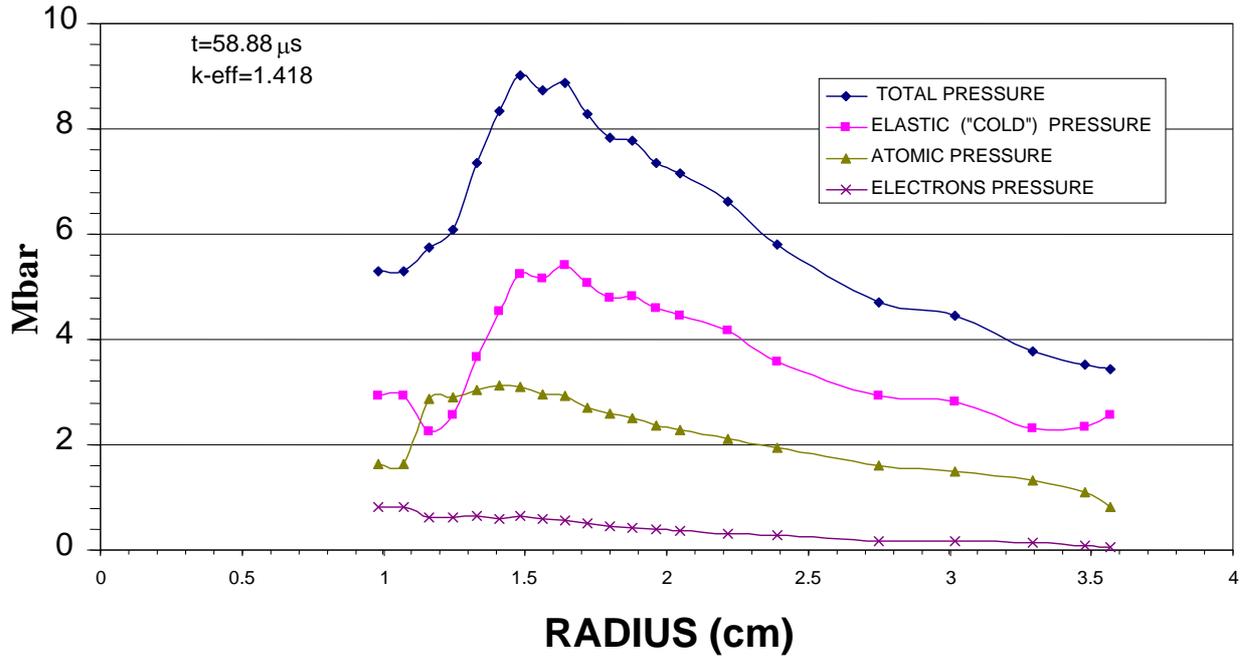

**Figure 21:** Contributions of the elastic (cold) pressure and the thermal pressures of atoms and electrons to the total pressure inside the Pu, at t=58.88 μs.

## 9. Conclusion

In this study, we determined the three-term equation of state (eq.(1)) for uranium and plutonium, appropriate for regimes in which these materials are submitted to strong shock compression (pressures in the range of megabars). For uranium, it is considered that its equation of state is accurate in all pressure ranges, as shown in the comparison with experimental data presented in section 6.

In the case of plutonium, the application of its equation of state for low pressures, where this element has different properties and allotropic phases, should be done with caution. For high pressures, the equation of state defined in this work may be applied both to the case of plutonium initially stabilized in delta-phase as to the case of plutonium at normal alpha-phase density. In the first case, as we considered an abrupt and instantaneous phase change to the alpha-phase (keeping the pressure negligible), it is not necessary to define any equation of state for delta-phase plutonium. This is due to the relatively low pressure at which the transition begins (around 6 kbar).

There is an important result with respect to the phase transitions that occur in the plutonium under shock compression. In the dynamic compression of alpha-phase plutonium, the increases of temperature in the plane shock wave is enough to lead it to the liquid state at P≅1.4 Mbar (Fig.16), thus before reaching the pressure corresponding to the transition to bcc phase, which occurs when plutonium is submitted only to the static pressure (see Fig.7). Now with regard to dynamic compression of plutonium stabilized in delta-phase, the much higher temperature reached in this case leads the plutonium directly from the alpha-phase to the liquid phase.

Finally, the equation of state defined here for the plutonium was used in the numerical simulation of the atomic bomb "Fat Man" (in a supposed configuration), with interesting results which are consistent with what is known about this that was the first implosion nuclear device designed in the early 40s, in the famous Manhattan Project.